\documentclass[final,5p,times,twocolumn]{elsarticle}

\usepackage{mathrsfs}
\usepackage{latexsym}
\usepackage{longtable}
\usepackage{graphicx}
\usepackage{epsfig}
\usepackage{natbib}
\usepackage{amsmath}
\usepackage{bbm}
\usepackage{amssymb}
\usepackage[usenames]{color}
\usepackage{CJK}
\usepackage{hyperref}
\usepackage{multirow}
\allowdisplaybreaks
\begin{document}

%\begin{CJK*}{GBK}{song}

\title{Evolution of proto-neutron stars with the hadron-quark phase transition}

\author[rvt]{Guo-yun Shao\corref{cor1}}\ead{shaogy@pku.edu.cn}
\cortext[cor1]{Corresponding author}
\address[rvt]{INFN-Laboratori Nazionali del Sud, Via S. Sofia 62, I-95123 Catania, Italy}

\begin{abstract}
The Poyakov-Nambu--Jona-Lasinio~(PNJL) model was developed recently,
which includes both the chiral dynamics and (de)confinement effect
and gives a good description of lattice QCD data. In this study we
use the PNJL model to describe the quark phase, and first use it to
study the evolution of  proto-neutron star~(PNS) with a hadron-quark phase transition.
Along the line of a PNS evolution, we take several snapshots of PNS
profiles, presenting the fractions of different species, the
equations of state~(EOS), and the mass-radius relations at different stages.
The calculation shows the mixed phase may exist during the whole evolving process,
and the onset density of quark phase decreases with the radiation of  neutrinos
in the heating stage. In the cooling stage, the EOS of the mixed phase softens
and the center density increases. In this process a part of nuclear matter
transforms to quark matter, which may lead to a PNS collapsing
into a black hole.

\end{abstract}

\maketitle

%\section{Introduction}
Since the matter in the core of compact stars are compressed to
densities of several times of saturated nuclear density, it is expected
that new degrees of freedom  will appear in the interior of these objects
\cite{Glendenning99,Menezes,Brown06,
Maruyama06,Shao09,Muto,Blaschke1999,Baldo03,
Weber2005,Alford20057,HShen081,Lattimer2007,Klahn2007}.
The hadron-quark phase transition is one of the most concerned topics
in modern physics related to heavy-ion collision experiment and compact star.
Because of the complication of  full calculation of Quantum Chromodynamics~(QCD)
and the lack of sufficient
knowledge about the nonpertubtive and (de)confinement effect,
it is difficult to apply full QCD calculation to describe the phase transition in
astrophysics. Therefore, in literatures the hadron-quark phase transition
related to compact star are usually described with the simplified,
phenomenological MIT bag model or the effective chiral
Nambu-Jona-Lasinio~(NJL) model~(e.g., \cite{
Blaschke1999, Baldo03, Weber2005, HShen081, Pagliara09, Buballa04, Yasutake09, Paoli10,Bombaci11}).

The NJL model with chiral dynamics is a prominent one in the application of
astrophysics, but it lacks the confinement mechanism, one  essential
characteristic of QCD. Recently, an improved version of the NJL model coupled to Polyakov loop
(PNJL) has been proposed~\cite{Fukushima04}. The PNJL model includes both the chiral dynamic and
(de)confinement effect, giving a good interpretation of
lattice QCD data~\cite{Ratti06,Costa10,Schaefer10,Herbst11,Kashiwa08,Abuki08,Fu08}.
We have recently studied  the hadron-quark phase transition relevant to heavy-ion collision
in the Hadron-PNJL model~\cite{Shao112}. The calculation
shows the color (de)confinement effect is
very important for the hadron-quark phase transition at finite density and temperature,
and improves greatly the results derived  from the Hadron-NJL model~\cite{Shao11}.
The PNJL model with a chemical potential dependent Poyakov effective potential
has also been used to describe cold neutron star without~\cite{Dexheimer10} and with a color
superconductivity structure~\cite{Blaschke10}. In~\cite{Fischer11},
Fischer \emph{et al}. discussed the evolution of core
collapse supernova with the PNJL model (with diquark interaction and isoscalar vector interaction)
and the possibility of the onset of deconfinement in core collapse supernova simulation.
In their study selected proton-to-baryon ratio $Y_p$ and the
Maxwell construction are taken for the PNJL model, which
gives a relatively narrow mixed phase than that with Gibbs construction.
In~\cite{Ohnishi11}, the authors studied the possibility to probe
the QCD critical endpoint during the dynamical black hole formation
from a gravitational collapse.  Several two-flavor quark models
with different parameters are used in their study, and the calculation
shows the Critical Point location of QCD has a strong
dependence on quark models and parameters.
A generalization with $s$ quark is needed to get more reliable
results for further investigation.

In this Letter we will firstly use the newly improved PNJL model to describe
the evolution of proto-neutron star from its birth with trapped neutrinos
to neutrino-free cold neutron star~(NS). The Gibbs criteria will be used to
determine the mixed phase under the isotropic constraint with
and without trapped neutrinos. The emphases are put on the evolution
of proto-neutron star with a hadron-quark phase transition,
the  particle distributions along baryon density, the EOSs and mass-radius
relations, as well as the star stability in different snapshots during the evolution.

A PNS forms after the gravitational collapse of the core
of massive star with the explosion of a supernova. At the beginning of the birth
of a PNS, the entropy per baryon is about one~($S\simeq1$) and the number of
leptons per baryon with trapped neutrino is approximate
0.4~($Y_{Le}=Y_e+Y_{\nu_e}\simeq0.4$). In the following
$10-20$ seconds, neutrinos escape from the star.
With the decrease of electron neutrino population, the star matter
is heated by the diffusing neutrinos, and the corresponding
entropy density increases, reaching to $S\simeq 2$ when $Y{_\nu{_e}} \simeq 0$.
Following the heating, the star begins cooling by radiating neutrino pairs of all flavors,
and finally a cold neutrons forms~\cite{Steiner00,Reddy98}.

Along the line of a PNS evolution to the formation of a cold neutron star, we
take several snapshots to study how the star evolves, especially with the appearance
of quark degrees of freedom. The snapshots are taken with the following
conditions ($ S=1,\,Y_{Le}=Y_e+Y_{\nu_e}= 0.4$),\,($S=1.5,\,Y_{Le}=Y_e+Y_{\nu_e}=0.3$),
($S=2,\,Y_{\nu_e}=0$) and ($S=0,\,Y_{\nu_e}= 0$), similar with that used
in \cite{Steiner00,Reddy98}. At each snapshot of a PNS evolution, we take an isentropic
approximation with which the temperature has a radial gradient in the star.
There are also studies related to the properties of PNS based
on isothermal approximation~(e.g., \cite{Burgio08,Yasutake09,Paoli10, Bombaci11} ).
%
%The aim of this study is to improve the calculation by incorporating recent
%developments in the description of quark matter and to perform a systemic
%study of the evolution of PNS star with a hadron-quark phase transition, particularly
%on the distribution of different particles, the EOSs and the mass-radius relations as well
%as the star stability during the evolution.

%\section{ neutron star matter}
For the star matter, the hadronic and quark phase are described by the non-linear Walecka
model and the PNJL model, respectively. In the
mixed phase between pure hadronic and quark matter,
the two phases are connected to each other by the Gibbs
conditions deduced from thermal, chemical and mechanical equilibriums.
For the hadron phase we use the Lagrangian given in \cite{Glendenning91} in which the interactions between
nucleons are mediated by $\sigma,\,\omega,\,\rho$ mesons, and the parameter set GM1 is used
in the calculation. The details can be found in Refs.~\cite{Glendenning91,Shao09}.

For the quark phase, we take recently developed three-flavor PNJL model with
the Lagrangian density
\begin{eqnarray}
\mathcal{L}_{q}&=&\bar{q}(i\gamma^{\mu}D_{\mu}-\hat{m}_{0})q+
G\sum_{k=0}^{8}\bigg[(\bar{q}\lambda_{k}q)^{2}+
(\bar{q}i\gamma_{5}\lambda_{k}q)^{2}\bigg]\nonumber\\
           &&-K\bigg[\texttt{det}_{f}(\bar{q}(1+\gamma_{5})q)+\texttt{det}_{f}
(\bar{q}(1-\gamma_{5})q)\bigg]\nonumber \\ \nonumber \\
&&-\mathcal{U}(\Phi[A],\bar{\Phi}[A],T),
\end{eqnarray}
where $q$ denotes the quark fields with three flavors, $u,\ d$, and
$s$, and three colors; $\hat{m}_{0}=\texttt{diag}(m_{u},\ m_{d},\
m_{s})$ in flavor space; $G$ and $K$ are the four-point and
six-point interacting constants, respectively. The four-point
interaction term in the Lagrangian keeps the $SU_{V}(3)\times
SU_{A}(3)\times U_{V}(1)\times U_{A}(1)$ symmetry, while the 't
Hooft six-point interaction term breaks the $U_{A}(1)$ symmetry.

The covariant derivative in the Lagrangian is defined as $D_\mu=\partial_\mu-iA_\mu$.
The gluon background field $A_\mu=\delta_\mu^0A_0$ is supposed to be homogeneous
and static, with  $A_0=g\mathcal{A}_0^\alpha \frac{\lambda^\alpha}{2}$, where
$\frac{\lambda^\alpha}{2}$ is $SU(3)$ color generators.
The effective potential $\mathcal{U}(\Phi[A],\bar{\Phi}[A],T)$ is expressed in terms of the traced Polyakov loop
$\Phi=(\mathrm{Tr}_c L)/N_C$ and its conjugate
$\bar{\Phi}=(\mathrm{Tr}_c L^\dag)/N_C$. The Polyakov loop $L$  is a matrix in color space
\begin{equation}
   L(\vec{x})=\mathcal{P} \mathrm{exp}\bigg[i\int_0^\beta d\tau A_4 (\vec{x},\tau)   \bigg],
\end{equation}
where $\beta=1/T$ is the inverse of temperature and $A_4=iA_0$.

Different effective potentials were
adopted in literatures~\cite{Ratti06,Robner07,Fukushima08,Dexheimer10}.
The modified chemical dependent one
\begin{eqnarray}
    \mathcal{U}&=&(a_0T^4+a_1\mu^4+a_2T^2\mu^2)\Phi^2\nonumber \\
               & &a_3T_0^4\,\mathrm{ln}(1-6\Phi^2+8\Phi^3-3\Phi^4)
\end{eqnarray}
was used in~\cite{Dexheimer10, Blaschke10} which is a simplification of
\begin{eqnarray}
    \mathcal{U}&=&(a_0T^4+a_1\mu^4+a_2T^2\mu^2)\bar{\Phi}\Phi\nonumber \\
               & &a_3T_0^4\,\mathrm{ln}\bigg[1-6\bar{\Phi}\Phi+4(\bar{\Phi}^3+\Phi^3)-3(\bar{\Phi}\Phi)^2\bigg]
\end{eqnarray}
because the difference between $\bar{\Phi}$ and $\Phi$ is smaller
at finite chemical potential and $\bar{\Phi}=\Phi$ at $\mu=0$. In the calculation we will use
the later one.
The related parameters, $a_0=-1.85,\,a_1=-1.44\times10^{-3},\,a_2=-0.08,\,a_3=-0.4$,
are still  taken from~\cite{Dexheimer10},
which can reproduce well the data obtained in lattice QCD calculation.

In the mean field approximation, quarks can be taken as free quasiparticles
with constituent masses $M_i$,
and the dynamical quark masses~(gap equations) are obtained as
\begin{equation}
M_{i}=m_{i}-4G\phi_i+2K\phi_j\phi_k\ \ \ \ \ \ (i\neq j\neq k),
\label{mass}
\end{equation}
where $\phi_i$ stands for quark condensate.
The thermodynamic potential of the PNJL model in the mean field level can be derived as
%expressed as
%\begin{eqnarray}
%\Omega&=&\mathcal{U}(\bar{\Phi}, \Phi, T)+2G\left({\phi_{u}}^{2}
%+{\phi_{d}}^{2}+{\phi_{s}}^{2}\right)-4K\phi_{u}\,\phi_{d}\,\phi_{s}\nonumber \\
%&&-T \sum_n\int \frac{\mathrm{d}^{3}p}{(2\pi)^{3}}\mathrm{Trln}\frac{S_i^{-1}(i\omega_n,\vec{p})}{T}.
%\end{eqnarray}
%Here $S_i^{-1}(p)=-(p\!\!\slash-M_i+\gamma_0(\mu_i-iA_4))$ is the inverse fermion propagator in the background field $A_4$,
%and the trace has to be taken in color, flavor, and Dirac space.
%After summing over the fermion  Matsubara frequencies, $p_0=i\omega_n=(2n+1)\pi T$,
%the thermodynamic potential can be written as
%\begin{widetext}
\begin{eqnarray}
\Omega&=&\mathcal{U}(\bar{\Phi}, \Phi, T)+2G\left({\phi_{u}}^{2}
+{\phi_{d}}^{2}+{\phi_{s}}^{2}\right)\nonumber \\
&&-4K\phi_{u}\,\phi_{d}\,\phi_{s}-2\int_\Lambda \frac{\mathrm{d}^{3}p}{(2\pi)^{3}}3(E_u+E_d+E_s) \nonumber \\
&&-2T \sum_{u,d,s}\int \frac{\mathrm{d}^{3}p}{(2\pi)^{3}} \mathrm{ln}\,\bigg[A(\bar{\Phi},\Phi,E_i-\mu_i,T)\bigg]\nonumber \\
&&-2T \sum_{u,d,s}\int \frac{\mathrm{d}^{3}p}{(2\pi)^{3}} \mathrm{ln}\,\bigg[\bar{A}(\bar{\Phi},\Phi,E_i+\mu_i,T)\bigg],
\end{eqnarray}
%\end{widetext}
where $A(\bar{\Phi},\Phi,E_i-\mu_i,T)=1+3\Phi e^{-(E_i-\mu_i)/T}+3\bar{\Phi} e^{-2(E_i-\mu_i)/T}+e^{-3(E_i-\mu_i)/T}$
and $\bar{A}(\bar{\Phi},\Phi,E_i+\mu_i,T)=1+3\bar{\Phi} e^{-(E_i+\mu_i)/T}+3\Phi e^{-2(E_i+\mu_i)/T}+e^{-3(E_i+\mu_i)/T}$.

The values of $\phi_u, \phi_d, \phi_s, \Phi$ and $\bar{\Phi}$ are determined by minimizing the thermodynamical
potential
\begin{equation}
\frac{\partial\Omega}{ \partial\phi_u}=\frac{\partial\Omega}{\partial \phi_d}=\frac{\partial\Omega}{\partial \phi_s}=\frac{\partial\Omega}{\partial \Phi}=\frac{\partial\Omega}{\partial \bar\Phi}=0.
\end{equation}
All the thermodynamic quantities relevant to the bulk properties of quark matter can be obtained from $\Omega$. Particularly, the pressure and entropy density
can be derived with $P=-(\Omega(T,\mu)-\Omega(0,0))$ and $S=-\partial \Omega / \partial T$, respectively.

As an effective model, the (P)NJL model is not
renormalizable, so a cut-off $\Lambda$ is implemented in 3-momentum
space for divergent integrations. The model parameters:
$\Lambda=603.2$ MeV, $G\Lambda^{2}=1.835$, $K\Lambda^{5}=12.36$,
$m_{u,d}=5.5$  and $m_{s}=140.7$ MeV, determined
by fitting $f_{\pi},\ M_{\pi},\ m_{K}$ and $\ m_{\eta}$ to their
experimental values~\cite{Rehberg95}, are used in the calculation.

The Gibbs criteria is usually implemented for a complicated system
with more than one conservation charge. The Gibbs conditions for
the mixed phase of hadron-quark phase transition in compact star are
\begin{eqnarray}
& &\mu_\alpha^H=\mu_\alpha^Q,\ \ \ \ \  \ T^H=T^Q,\ \ \ \ \ \ P^H=P^Q,
\end{eqnarray}
where $\mu_\alpha$ are usually chosen with $\mu_n$ and $\mu_e$.
Under the $\beta$ equilibrium with trapped neutrino, the chemical potential
of other particles including all baryons, quarks, and leptons can be derived by
\begin{equation}
    \mu_i=b_i\mu_n-q_i\mu_e+q_i\mu_{\nu_e},
\end{equation}
where $b_i$ and $q_i$ are the baryon number and electric charge
number of particle species $i$, respectively. For the matter with trapped
electron neutrinos,
$Y_{L\mu}=(Y_\mu+Y_{\nu_\mu})\simeq 0$, we do not need to
consider the contribution from muon and muon neutrino~\cite{Steiner00,Reddy98}. For the
neutrino-free matter~($\mu_{\nu_e}=0$), both electrons and muons are included in the calculation.

The baryon number density and energy density in the mixed phase are
composed of two parts with the following combinations
\begin{equation}
    \rho=(1-\chi)\rho_B^H+\chi\rho_B^Q,
\end{equation}
and
\begin{equation}
    \varepsilon=(1-\chi)\varepsilon^H+\chi\varepsilon^Q,
\end{equation}
where $\chi$ is the volume fraction of quark matter. And the electric neutrality is fulfilled globally
with
\begin{equation}
    q_{total}=(1-\chi)\sum_{i=B,l}q^i \rho_i+\chi \sum_{i=q,l}q^i \rho_i=0.
\end{equation}

%\section{Summary}

\begin{figure}[htbp]
\begin{center}
\includegraphics[scale=0.3]{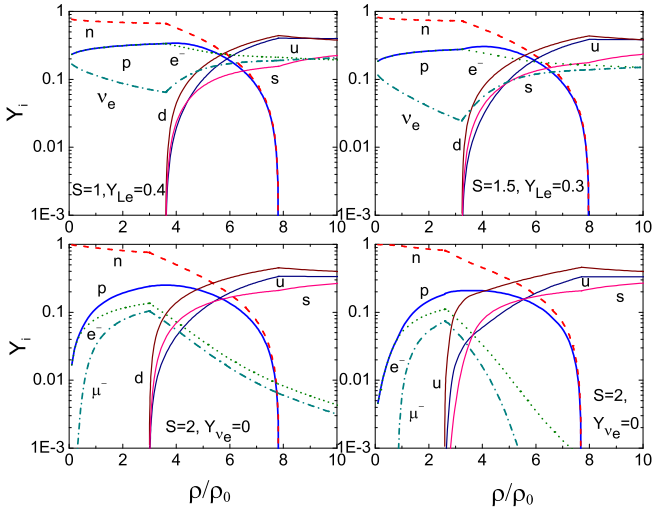}
\caption{\label{fig:GM1-Xi}Relative fractions of different species as
functions of baryon density at several snapshots of a PNS evolution. The upper~(lower) panels
are the results with~(without) trapped neutrinos}
\end{center}
\end{figure}
In Fig.~\ref{fig:GM1-Xi}, we display  the relative fractions
of different species as functions
of baryon density in several snapshots along the
evolution of a proto-neutron star, from ($ S=1,\,Y_{Le}=Y_e+Y_{\nu_e}= 0.4$),\,
($S=1.5,\,Y_{Le}=Y_e+Y_{\nu_e}=0.3$),\,($S=2,\,Y_{\nu_e}=0$)
to cold neutron star with ($S=0,\,Y_{\nu_e}= 0$).
Comparing the upper panels with trapped neutrino with the
lower panels without trapped neutrino, we find that
the fraction of trapped neutrino affects the proton-neutron ratio
$Y_p/Y_n$ at lower density before the appearance of quarks.
For the hot PNS matter, $Y_p/Y_n$ with rich $\nu_e$
is larger than that with poor $\nu_e$. According to the Pauli principle,
the smaller $Y_p/Y_n$ will
excite more neutrons to occupy higher energy levels,
leading to a stiffer equation of state.

Another point we stress is that the onset density of quark phase decreases
with the escape of trapped neutrinos, i.e., trapped neutrinos delay
the hadron-quark phase transition to a higher density. For the neutrino-trapped
matter, the fraction of neutrino is enhanced with the appearance of
quarks, which also affects the neutrino opacity. For neutrino-free cases,
the lepton population in cold neutron star matter ($S=0,\,Y_{\nu_e}=0$)
is smaller than that of PNS ($S=2,\,Y_{\nu_e}=0$), especially after
the appearance of quarks at high density.

The largest center densities of proto-neutron stars at the first three snapshots
taken above are $\rho_c^{}=5.23,\,
5.00,$ and $4.54\,\rho_0$, respectively. The center density decreases with the radiation of neutrinos,
because the star expands when the inner matter is heated by the escaped neutrinos.
When neutrinos are free, the PNS begins cooling by radiating neutrino pairs of all flavor
and the star shrinks until the formation of cold neutron star.
During the cooling stage, the center density of the star increases. The center density
of cold neutron star is $\rho_c^{}=4.88\,\rho_0^{}$. The variation of the center densities
during a PNS evolution is easier to understand by
combining the mass-radius relations that will be given latterly.

\begin{figure}[htbp]
\begin{center}
\includegraphics[scale=0.3]{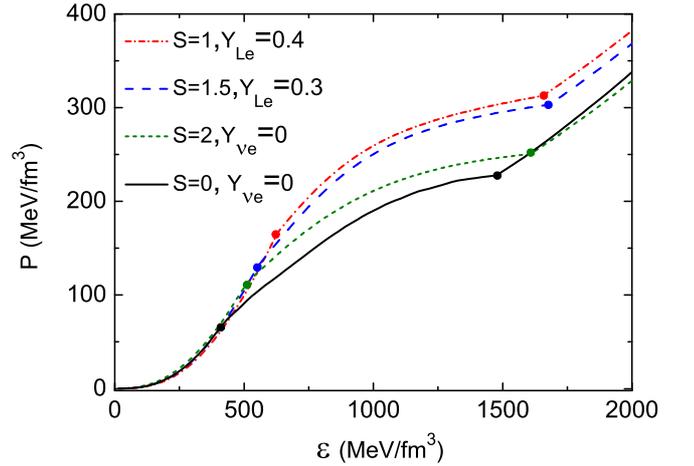}
\caption{\label{fig:GM1-EOS}EOSs of PNS and NS matter at several stages of the star evolution.
The dots mark the range of the mixed phase.}
\end{center}
\end{figure}

We present the equations of state of the star matter
at different evolving stages in Fig.~\ref{fig:GM1-EOS}.
The dots mark the ranges of the mixed phases. For the PNS matter
at low density before the onset of quarks, the EOS becomes more and
more stiffer with the decrease of lepton fraction and the increase
of entropy density. This mainly attributes
to the decreased  $Y_p/Y_n$, as shown in Fig.~\ref{fig:GM1-Xi},
which excites more neutrons to higher energy states.
In contrast, the EOS of the mixed phase with a larger lepton fraction
is much stiffer. This is because the pure quark phase has a stiffer EOS, but the
corresponding hadron phase with the same lepton fraction and entropy density has
a softer one. To fulfill the Gibbs condition of the chemical and mechanical equilibrium,
the phase transition can only take place at relatively larger energy density.
For the case with a stiffer hadronic EOS at lower density, the
Gibbs condition can be realized at lower energy density to drive the hadron-quark
phase transition. The similar results have been obtained when the NJL model is taken~\cite{Steiner00}.
The cold neutron star matter at lower density has an EOS between the initial conditions and the
end of heating stage, and a softest one for the mixed phase after quarks appearing.

Comparing the results obtained in the NJL model in~\cite{Steiner00}, we find the pressure of the
mixed phase of a PNS given by the PNJL model is larger than that of NJL model.
This reflects that the confinement effect~(gluon field) is important
at finite temperature. With
the PNJL model, the pressure of quark matter at finite temperature
are much smaller than that of NJL model, therefore the phase transition
can only take place at relatively larger density to assure
the quark-phase pressure can match that of the hadron phase. The details can be
found in refs.~\cite{Shao112, Shao11} where the hadron-quark
phase transition for symmetric and asymmetric matter related to heavy-ion collision experiment
has been investigated in the Hadron-(P)NJL model, and the same results were obtained.

The above illustration also explains the inverse of pressure of the mixed phase
in the cases of ($S=2,\,Y_{\nu_e}=0$) and ($S=0,\,Y_{\nu_e}=0$) as given
by the NJL model in~\cite{Steiner00}.
All the features of equations of state will be reflected in the mass-radius relations.

\begin{figure}[htbp]
\begin{center}
\includegraphics[scale=0.3]{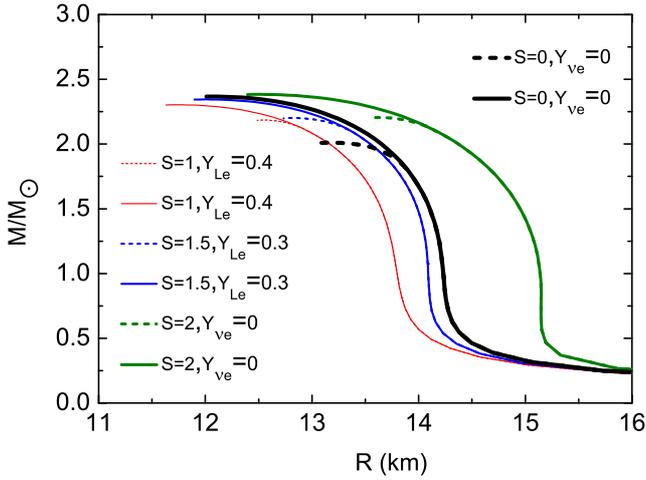}
\caption{\label{fig:GM1-RM}Mass-radius relations of PNS and NS at several snapshots
along a PNS evolution.
The solid curves are the results without quarks and the dash curves are
the results with hadron-quark phase transition.}
\end{center}
\end{figure}
Finally, in Fig.~\ref{fig:GM1-RM} we plot the mass-radius relations of several snapshots
taken above along the evolution a PNS. This figure gives us
a more intuitive picture about the PNS evolution with a
hadron-quark phase transition. Firstly, in the heating stage the
PNS expands  with the decrease of lepton fraction~($Y_{Le}$)
and the increase of entropy density~($S$).
Simultaneously, the center density  decreases with the
expansion of the star.
When neutrinos are free, the star begins cooling, and then the star shrinks,
leading to the center density increasing again.
In this stage for a star with quarks, a
part of nuclear matter transforms to quark matter and the EOS
becomes softer in the mixed phase. The star may
collapse into a black hole in the cooling process if the EOS in the core cannot
resist the gravity. This point is different from the result derived in the NJL model.

The radio timing observations of the binary
millisecond pulsar J1614-2230 with a
strong general relativistic Shapiro delay signature, implies
that the pulsar mass is $1.97\pm0.04\,\mathrm{M}_\odot$~\cite{Demorest10}. The
discovery of this massive pulsar rules out many
soft equations of state. With the improved quark model, our calculation shows
the maximum mass of cold neutron star with deconfined quarks
is slightly larger than $2\,\mathrm{M}_\odot$.
Because the quark model parameters are determined by experiments and lattice QCD
simulation, a stiffer hadronic EOS is needed to fulfill this constraint.
On the other hand, If a vector interaction is included
for quark matter, the EOS will be stiffer and the maximum mass of hybrid star will be improved.

In summary, we have studied the evolution of PNS with a hadron-quark phase transition with
a more reliable quark model including both chiral dynamics and (de)confinement effect.
The calculation shows the quark phase may exist in the whole
process of a PNS evolution. The trapped neutrinos affect greatly the ration
of $Y_p/Y_n$ and the EOS at low density before quarks appearing.
In the heating stage, with the deleptonization and
the increase of entropy density, the PNS expands and the corresponding center density
decreases. In contrast, in the cooling stage, the PNS shrinks and the center density increases.
During this process a part of nuclear matter transforms to quark matter, and the
PNS may collapse into a black hole  if the EOS in the core is not stiff enough.

In this study, only the $\bar{q}q$ interaction is considered for quark phase
because the relevant model parameters can be fixed with experiments
and lattice QCD simulation. In analogy to BCS theory,
color superconductivity in low-temperature and high-density QCD
matter may appear and there may exist rich phase diagram.
The coupling constant of this interaction channel affects the equation
of state of quark matter and the onset density of quark phase in compact star.
One question is that the coupling parameter  cannot be fixed from
heavy-ion experiment or  lattice QCD  simulation.
On the other hand,  the (isoscalar) vector interaction channel
has been included in some studies.  Such an interaction
reduces the effective quark chemical potential but contributes
to the pressure of quark matter. This interaction stiffens
the EOS of quark matter and increases the maximum mass of a hybrid star.
Compared with the hadron Walecka  model, the  (isoscalar) vector
interaction in quark matter plays the role corresponding to the $\omega$ meson.
One drawback is that the coupling constant is
usually taken as a free parameter and its strength affects greatly the
Critical End Point of chiral symmetry restoration relevant to heavy-ion collision experiment.
Besides,  the isovector vector interaction (corresponding to the role
of $\rho$ meson in hadron phase and influencing the symmetry energy
of quark matter) can be also introduced  for  asymmetric quark matter.
For these  interaction channels as well as hyperon degrees of freedom,
the difficulty is the uncertainties of the relevant coupling parameters with
the lack of experiment data, so we temporarily omit these interactions in this study.
And a systematic investigation on these problems is
in progress as a further study.

%In this study we temporarily do not  consider hyperons due to the uncertainty of hyperon-meson
%coupling constants, but a systematic analysis with hyperon degrees of freedom will be given
%in the future.


\begin{thebibliography}{99}

\bibitem{Glendenning99}N. K. Glendenning  and J. Schaffner-Bielich,
         Phys. Rev. Lett. 81 (1998) 4564; Phys. Rev. C 60  (1999) 025803;
         J. Schaffner and I. N. Mishustin, Phys. Rev. C  53 (1996) 1416;
         J. Schaffner-Bielich, Nucl. Phys. A 804 (2008) 309.
\bibitem{Menezes}D. P. Menezes, P. K. Panda, and C. Provid\^{e}ncia, Phys. Rev. C 72 (2005) 035802.
\bibitem{Brown06}G. E. Brown, C. H. Lee, H. J. Park, and M. Rho, Phys. Rev. Lett. 96 (2006) 062303.
\bibitem{Maruyama06}T. Maruyama, T. Tatsumi, D. N. Voskresensky,
         T. Tanigawa, T. Endo, and S. Chiba, Phys. Rev. C 73 (2006) 035802.
\bibitem{Shao09}G. Y. Shao and Y. X. Liu, Phys. Rev. C 79 (2009) 025804;
        Phys. Rev. C  82  (2010) 055801; Phys. Lett. B   682 (2009) 171.
\bibitem{Muto}T. Muto, Nucl. Phys. A  754 (2005) 350; Phys. Rev. C 77 (2008) 015810.
\bibitem{Blaschke1999} D. B. Blaschke, H. Grigorian, G. Poghosyan, C. D. Roberts, and S. Schmidt,
        Phys. Lett. B  450 (1999) 207; D. Blaschke, {\it et al}, Phys. Rev. D 72 (2005) 065020.
\bibitem{Baldo03}M. Baldo, {\it et al}., Phys. Lett. B 562 (2003) 153.
\bibitem{Weber2005}F. Weber, Prog. Part. Nucl. Phys.  54 (2005) 193 .
\bibitem{Alford20057} M. Alford, M. Braby, M. Paris, and S. Reddy,  Astrophys. J. 629 (2005) 969;
       M. Alford, D. B. Blaschke, A. Drago, T. Kl\"{a}hn, G. Pagliara, and J. Shaffner-Bielich, Nature 445 (2007) E7.
\bibitem{HShen081} F. Yang  and H. Shen, Phys. Rev. C 77 (2008) 025801.
\bibitem{Lattimer2007}J. M. Lattimer and M. Prakash, Phys. Rep. 442 (2007) 109.
\bibitem{Klahn2007}T. Kl\"{a}hn, D. Blaschke, F. Sandin et al., Phys. Lett. B 654 (2007) 170.
\bibitem{Pagliara09}G. Pagliara, M. Hempel, and J. Schaffner-Bielich, Phys. Rev. Lett. 103 (2009) 171102;
        K. Schertler and S. Leupold, and J. Schaffner-Bielich, Phys. Rev. C 60 (1999) 025801.
\bibitem{Burgio08}G. F. Burgio and S. Plumari, Phys. Rev. D 77 (2008) 085022;
O. E. Nicotra, M. Baldo, G. F. Burgio, and H. J Schulze, Phys. Rev. D 74 (2006) 123001.
\bibitem{Buballa04}M. Buballa, F. Neumann, M. Oertel, and I. Shovkovy, Phys. Lett. B 595 (2004) 36.
\bibitem{Yasutake09}N. Yasutake and K. Kashiwa, Phys. Rev. D  79 (2009) 043012.
\bibitem{Paoli10}M. G. Paoli and D. P. Menezes, Eur. Phys. J. A 46 (2010) 413; D. P. Menezes and C. Provid\^{e}cia, Phys. Rev. C 68 (2003) 035804.
\bibitem{Bombaci11}I. Bombaci, D. Logoteta, C. Provid\^{e}ncia, and I. Vida\~{n}a, Astro. Astrophys. 528 (2011) A71.
\bibitem{Fukushima04}K. Fukushima, Nucl. Phys. B    591 (2004) 277.
\bibitem{Ratti06}C. Ratti, M. A. Thaler, W. Weise, Phys. Rev. D 73 (2006) 014019.
\bibitem{Costa10} P. Costa, M. C. Ruivo, C. A. de Sousa, and H. Hansen, Symmetry 2(3) (2010) 1338.
\bibitem{Schaefer10}B-J. Schaefer, M. Wagner, J. Wambach, Phys. Rev. D 81 (2010) 074013.
\bibitem{Herbst11} T. K. Herbst, J. M. Pawlowski, B-J. Schaefer, Phys. Lett. B  696 (2011) 58.
\bibitem{Kashiwa08}K. Kashiwa, H. Kouno, M. Matsuzaki,  and M. Yahiro, Nucl. Phys. B  662 (2008) 26.
\bibitem{Abuki08}H. Abuki, R. Anglani, R. Gatto, G. Nardulli and M. Ruggieri,
        Phys. Rev. D  78 (2008) 034034.
\bibitem{Fu08}W. J. Fu, Z. Zhang, Y. X. Liu, Phys. Rev. D 77 (2008) 014006.
\bibitem{Shao112}G. Y. Shao, M. Di Toro, V. Greco, M. Colonna, S. Plumari, B. Liu, and Y. X. Liu,
	Phys. Rev. D 84  (2011) 034028.
\bibitem{Shao11}G. Y. Shao, M. Di Toro, B. Liu, M. Colonna, V. Greco, Y. X. Liu, and S. Plumari,
Phys. Rev. D 83 (2011) 094033.
\bibitem{Dexheimer10}V. A. Dexheimer and S. Schramm, Phys. Rev. C 81 (2010) 045201;
Nucl. Phys. B 199 (2010) 319; R. Negreiros, V. A. Dexheimer, and S. Schramm,
Phys. Rev. C 82 (2010) 035803.
\bibitem{Blaschke10}D. Blaschke, J. Berdermann and R. {\L}astowiecki,
Prog. Theor. Phys. Suppl 186 (2010) 81.
\bibitem{Fischer11}T. Fischer et al., arxiv:1103.3004v2.
\bibitem{Ohnishi11}A. Ohnishi et al., arxiv:1102.3753v1.
\bibitem{Steiner00}A. W. Steiner, M. Praskash, J. M. Lattimer, Phys. Lett. B 486 (2000) 239.
\bibitem{Reddy98}S. Reddy, M. Praskash, J. M. Lattimer, Phys. Rev. D 58 (1998) 013009.
\bibitem{Glendenning91}N. K. Glendenning and S. A. Moszkowski, Phys. Rev. Lett. 67
(1991) 2414;  N. K. Glendenning, {\it Compact Stars} (Springer-Verlag, Berlin, 2000).
\bibitem{Robner07} S. R\"{o}{\ss}ner, C. Ratti, and W. Weise, Phys. Rev. D 75 (2007) 034007.
\bibitem{Fukushima08}K. Fukushima, Phys. Rev. D 77 (2008) 114028.
\bibitem{Rehberg95}P. Rehberg, S. P. Klevansky, and J. H\"{u}fner, Phys. Rev. C 53 (1995) 410.
\bibitem{Demorest10} P. B. Demorest, T. Pennucci, S. M. Ransom, M. S. E.
Roberts, and J. W. T. Hessels, Nature 467 (2010) 1081.
 
\end{thebibliography}
\end{document}